\documentclass[12pt]{article}
\usepackage{epsfig}
\usepackage[hang,bf,small]{caption}
\DeclareGraphicsExtensions{.eps.gz,.eps,.ps,.ps.gz}
\parskip=\itemsep               %?

\newcommand{\beq}{\begin{equation}}
\newcommand{\eeq}{\end{equation}}
\newcommand{\beqar}[1]{\begin{eqnarray}\label{#1}}
\newcommand{\eeqar}{\end{eqnarray}}

\newcommand{\si}{\sigma}

\newcommand{\as}{\alpha_S}

\def\eq#1{{Eq.~(\ref{#1})}}

\def\plb#1#2#3{    {\it Phys. Lett. }{\bf B#1} (19#2) #3}
\def\prd#1#2#3{    {\it Phys. Rev. }{\bf D#1} (19#2) #3}

\def\zpc#1#2#3{    {\it Z. Phys. }{\bf C#1} (19#2) #3}

\relax
\begin{document}
\title{
{ \Large \bf  Non-Linear Evolution and} \\
{\Large \bf   High Energy Diffractive Production}}
\author{
{\large  ~ E.~Levin\thanks{e-mail: 
leving@post.tau.ac.il}~~$\mathbf{{}^{a),b)}}$
 \,~\,and\,\,~
M. ~Lublinsky\thanks{e-mail: 
mal@techunix.technion.ac.il}~~$\mathbf{{}^{c),d)}}$}\\[2.5ex]
 {\it ${}^{a)}$ \small HEP Department}\\
{\it \small School of Physics and Astronomy}\\
{\it \small Raymond and Beverly Sackler Faculty of Exact Science}\\
{\it\small Tel Aviv University, Tel Aviv 69978, ISRAEL}\\[2.5ex]
{\it ${}^{b)}$ \small DESY Theory Group,}\\
{ \it\small D-22602, Hamburg, GERMANY}\\[2.5ex]
{\it ${}^{c)}$ \small Department of Physics}\\
{\it \small Technion -- Israel Institute of   Technology}\\
{\it\small  Haifa 32000, ISRAEL}\\[2.5ex]
{\it ${}^{d)}$ \small  II. Institut f\"{u}r Theoretische Physik,  Universit\"{a}t Hamburg}\\
{\it\small Luruper Chaussee 149, 
       22761 Hamburg, GERMANY  }\\[1.5ex]
}

\maketitle
\thispagestyle{empty}
                      
\begin{abstract} 
The ratio of the diffractive production to the total cross section in DIS
is computed as a function of the produced mass. The analysis is based on the solution to the
non-linear  evolution equation for the diffraction  dissociation in DIS.

The obtained ratios almost do not depend on the central mass
energy in agreement with the HERA experimental data. This independence is argued
to be a consequence of the scaling phenomena displayed by the cross sections.

As a weakness point a significant discrepancy between the data and the obtained
results is found in the absolute values of the ratios. Several explanatory 
reasons are discussed.

 \end{abstract}
\thispagestyle{empty}
\begin{flushright}
\vspace{-21.5cm}
DESY-01-124 \\
TAUP - 2689 - 2001 \\
\today
\end{flushright}   
\newpage
\setcounter{page}{1}

\section{Introduction}
\setcounter{equation}{0}

\begin{figure}[htbp]
\begin{center}
\epsfig{file=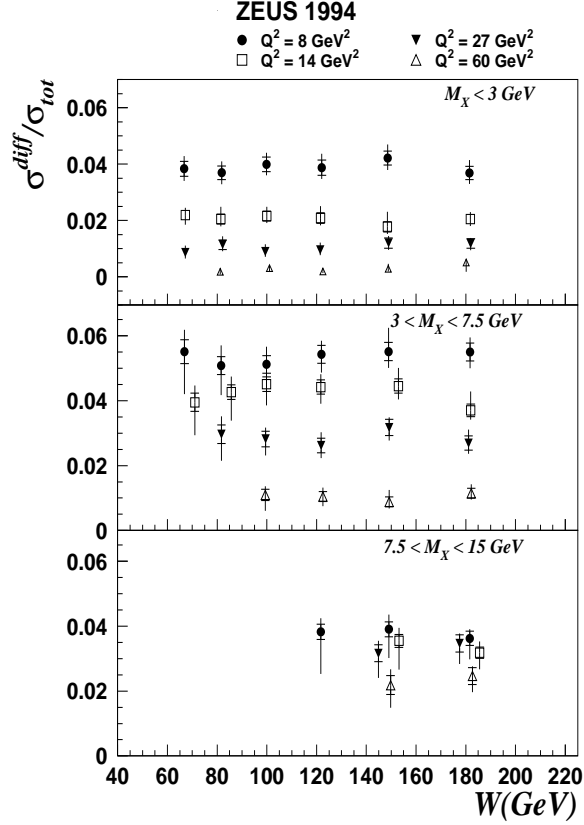,width=90mm,height=120mm}
\caption{\it Experimental data for the ratio
$\sigma_{diff}/\sigma_{tot}$ taken from Ref. \protect\cite{ZEUSDATA}. }
\label{fig1}
\end{center}
\end{figure}

One of the most intriguing  experimental
observations in HERA  is in the energy independence of
  the ratio between the  cross  section of single 
diffractive dissociation and the total DIS cross section  \cite{ZEUSDATA} 
(see Fig.\ref{fig1}).
 The widely used saturation model of Golec-Biernat and 
Wusthoff (GW) quite successfully reproduces this data \cite{GW}.
It was conjectured by Kovchegov and McLerran that 
the effects of  the parton density  saturation 
 occuring at high energies are responsible for this independence \cite{KM}.
Using the unitarity constraint they related the diffraction cross section 
($\sigma_{diff}$) to the total cross section ($\sigma_{tot}$) 
in DIS of $q\bar q$ pair with a target
\beq \label{KMF}
R\,\,
\equiv\,\,\frac{\sigma_{diff}}{\sigma_{tot}}\,\,=\,\,\frac{\int\,d^2\,b \int
\,dz \int \,d^2  r_{\perp}
P^{\gamma^*}(z,r_{\perp};Q^2)\, N^2(r_{\perp},x;b)}
{ 2\,  \int\,d^2 b \,\int\,dz
\,\int\,d^2\,r_{\perp}\,\,P^{\gamma^*}(z,r_{\perp};Q^2)\,\,
N ( r_{\perp},x;b)
 }\,\,.
\eeq
The function  $N(r_{\perp},x; b) = Im
\,a^{el}_{\rm dipole}(r_{\perp},x; b)$, where $a^{el}_{\rm dipole}$ is 
the amplitude
of the elastic scattering for the dipole of the size $r_{\perp}$ and 
rapidity $Y\equiv\ln (1/x)$ 
 scattered at  impact parameter  $b$. The Bjorken $x$ is related to the
central mass energy $W$ via $x=Q^2/W^2$.
$P^{\gamma^*}(z,r_{\perp};Q^2)$ is the probability to find a 
quark-antiquark pair with the size $r_{\perp}$ inside the virtual photon 
\cite{MU90,WF}:
\begin{eqnarray}
P^{\gamma^*}(z,r_{\perp};Q^2)&=&\frac{\alpha_{em} N_c}{2
\pi^2}
\,\sum_f \,Z^2_f \sum_{\lambda_1,\lambda_2}\,\{\, | \Psi_T |^2\,\,+\,\,|
\Psi_L|^2 \,\}\,\,\label{PROBPH}\\
&=&\frac{\alpha_{em} N_c}{2 \pi^2}
\sum_f Z^2_f \,\{\,( z^2 + ( 1 - z )^2 )a^2 K^2_1( a\,r_{\perp}
)\,+\,4\,Q^2\,z^2( 1 - z )^2 K^2_0(
a\,r_{\perp})\,\},\nonumber
\end{eqnarray}
where in the quark massless limit $a^2 = z(1-z)Q^2$ and $\Psi_{T,L}$ stand for the
$q\bar q$ wave functions of transversely and longitudinally polarized photons.

For the amplitude $N$ 
a non-linear evolution equation was derived  
\cite{GLR,MUQI,MU94,BA,KO,Braun,ILM}. 
This equation  has been studied both 
analytically \cite{ILM,LT} and numerically \cite{Braun,LGLM,LL,Braun2}. 

Even with inclusion of an extra gluon emission  Eq. (\ref{KMF})
 fails to describe correctly the experimental data of Fig. \ref{fig1}
\cite{GLMDD,KOP,KOVN,KOV}. However, \eq{KMF} can be used as  initial 
condition to a further evolution.

Similarly to the total cross section 
we introduce  the cross section for diffractive production  with the 
 rapidity gap larger than given $Y_0\equiv\ln (1/x_0)$:
\beq
\label{F2D}
\si_{diff}(x,x_0,Q^2)\,\,\,=\,\,\int\,d^2b\,\,\int\,\,d^2 r_{\perp} \int \,d
z\,\,P^{\gamma^*}(z,r_{\perp};Q^2)
 \,\,N^D(r_{\perp},x,x_0;b)\,.
\eeq

The function $N^D$ is the amplitude
of the diffractive production induced by the dipole 
with the size $r_{\perp}$  and rapidity
gap larger than given ($Y_0$). 
The minimal rapidity gap $Y_0$ can be kinematically
related to the maximal diffractively  produced mass $x_0=(Q^2+M^2)/W^2$. 
The amplitude $N^D$ is a subject to 
a non-linear evolution equation derived for the diffraction dissociation 
processes in Ref. \cite{LK} and recently rederived in Ref. \cite{KOVN}:

\begin{eqnarray} 
   N^D({\mathbf{x_{01}}},Y,Y_0;b)  =  N^2({\mathbf{x_{01}}},Y_0;b)\, 
{\rm e}^{-\frac{4
C_F\,\as}{\pi} \,\ln\left( \frac{{\mathbf{x_{01}}}}{\rho}\right)(Y-Y_0)}\,
+\frac{C_F\,\as}{\pi^2}\int_{Y_0}^Y dy \,  {\rm e}^{-\frac{4
C_F\,\as}{\pi} \,\ln\left( \frac{{\mathbf{x_{01}}}}{\rho}\right)(Y-y)} \times  \nonumber \\
\nonumber \\
 \int_{\rho} \, d^2 {\mathbf{x_{2}}}  
\frac{{\mathbf{x^2_{01}}}}{{\mathbf{x^2_{02}}}\,
{\mathbf{x^2_{12}}}} 
[\,2\,  N^D({\mathbf{x_{02}}},y,Y_0;{ \mathbf{ b-
\frac{1}{2}
x_{12}}})
+  N^D({\mathbf{x_{02}}},y,Y_0;{ \mathbf{ b - \frac{1}{2}
x_{12}}})  N^D({\mathbf{x_{12}}},y,Y_0;{ \mathbf{ b- \frac{1}{2}
x_{02}}}) \nonumber \\  \label{DDEQ} \\
- 4 \, N^D({\mathbf{x_{02}}},y,Y_0;{ \mathbf{ b - \frac{1}{2}
x_{12}}})  N({\mathbf{x_{12}}},y;{ \mathbf{ b- \frac{1}{2}
x_{02}}})+2\, N({\mathbf{x_{02}}},y;{ \mathbf{ b -
\frac{1}{2}
x_{12}}})  N({\mathbf{x_{12}}},y;{ \mathbf{ b- \frac{1}{2}
x_{02}}})
]\,. \nonumber 
\end{eqnarray}
The equation (\ref{DDEQ}) describes a diffraction process initiated by  dipole of the size
$  \mathbf{x_{01}}$ which subsequently dissociates to two dipoles with the sizes 
$  \mathbf{x_{02}}$ and $  \mathbf{x_{12}}$.
The rapidity  $Y$ is defined as $Y=\ln (1/x)$.
First numerical solution of this equation   was recently
obtained in Ref. \cite{LL1}.
At the energy equal to the minimal
energy gap diffraction is purely given 
by the elastic scattering as it is stated in \eq{KMF}:
\beq\label{iniDD}
N^D(r_\perp,x_0,x_0;b)\,=\,N^2(r_\perp,x_0;b)\,.
\eeq

In the present letter we compute the ratio $\si_{diff}/\si_{tot}$ in mass bins.
For the function $N$ and $N^D$ we use the numerical solutions
obtained in Refs. \cite{LGLM,LL1}.

The letter is organized as follows. In the next section (2) we compute the 
$\si_{diff}/\si_{tot}$ ratio. To this goal we first study the 
$b$-dependence of the amplitude $N^D$. Discussion of the results is presented
in section 3. We conclude in the last section (4).

\section{$\si_{diff}/\si_{tot}$}
 
\begin{figure}[htbp]
\begin{tabular}{c c c}
 \epsfig{file=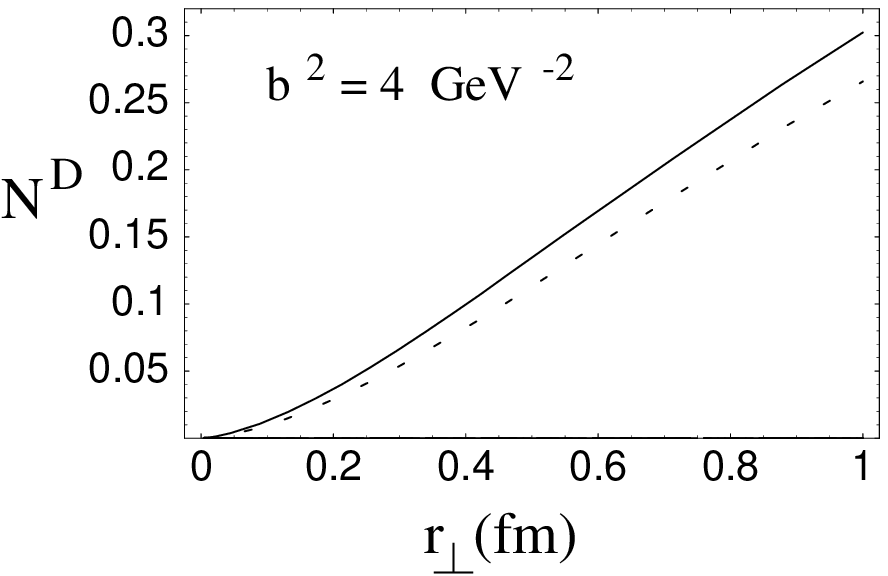,width=54mm, height=45mm}&
\epsfig{file=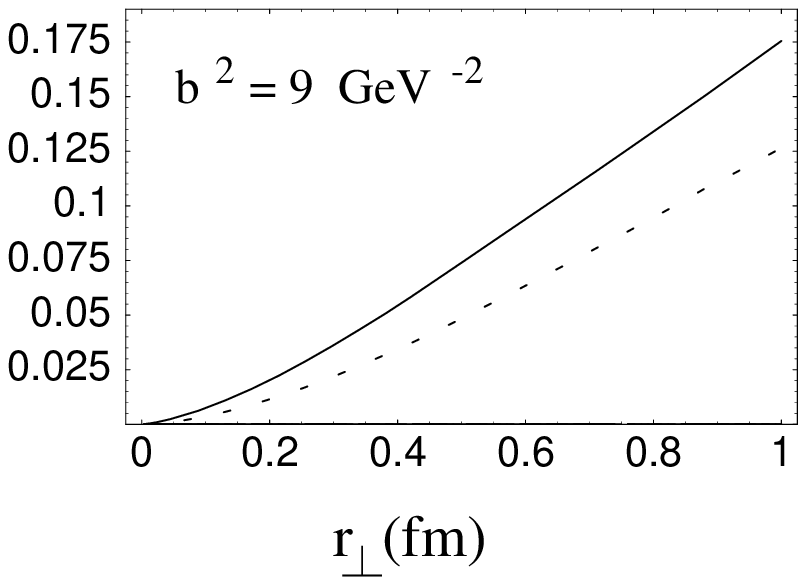,width=50mm, height=45mm}&
 \epsfig{file=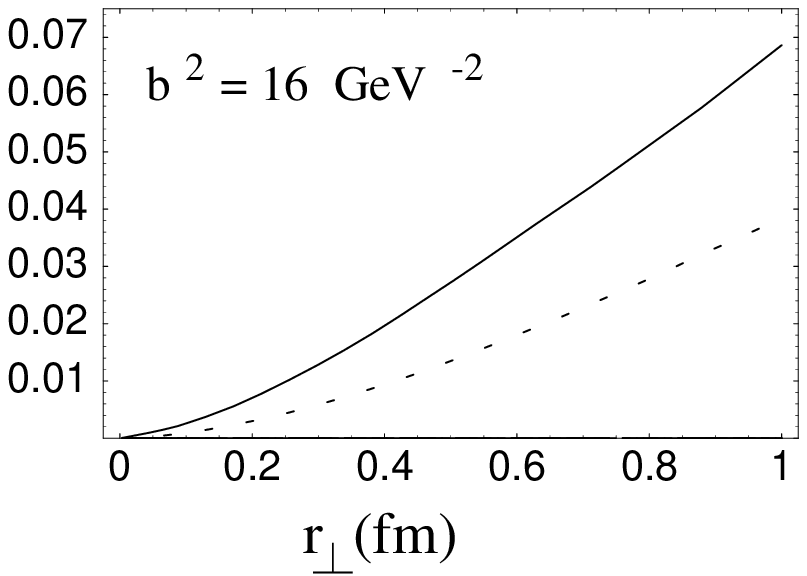,width=50mm, height=45mm}\\
\end{tabular}
  \caption[]{\it The comparison between the anzatz (\ref{NDb})(dashed line) 
and the true $b$-dependence (solid line). The curves are plotted as a 
function of distance  at fixed $x=10^{-3}$.}
\label{bdep}
\end{figure}

We assume the following
$b$-dependence of $N^D$:
\beq
\label{NDb}
 N^D(r_\perp,x,x_0; b)\,=\, (1\,-\,e^{-\kappa^D(x,x_0,r_\perp)\, S(b)})^2,
\eeq
with 
\beq
\label{kappaD}
\kappa^D(x,x_0,r_\perp)\,=\,-\,\ln(1\,-\,\sqrt{\tilde N^D(r_\perp,x,x_0)}).
\eeq
$\tilde N^D(r_\perp,x,x_0)$ computed in Ref. \cite{LL1} 
represents a solution of the 
same  equation (\ref{DDEQ}) but with no dependence on the forth variable. 
The initial conditions for the function 
$\tilde N^D(r_\perp,x,x_0)$ are set at $b=0$. In order to estimate the accuracy
of the anzatz (\ref{NDb}) the non-linear equation (\ref{DDEQ}) was solved
for several values of $b$ with the only assumption $r_\perp \ll b$. The 
comparison with the anzatz is shown in Fig. \ref{bdep}.

The anzatz  (\ref{NDb})  underestimates significantly
 the correct $b$-dependence
of the amplitude and the mismatch grows with $b$. Similar underestimation
was obtained for the function $N$ in Ref. \cite{LGLM} and it can be naturally
explained \cite{LL}. It is important to note, however, that the  mismatch
of the function $N^D$ is significantly larger than the one of the function 
$N$. In the final computation of the ratio this fact  leads to  
underestimation of the ratio especially for smaller $Q^2$.

\begin{figure}[htbp]
\begin{tabular}{c c c}
 \epsfig{file=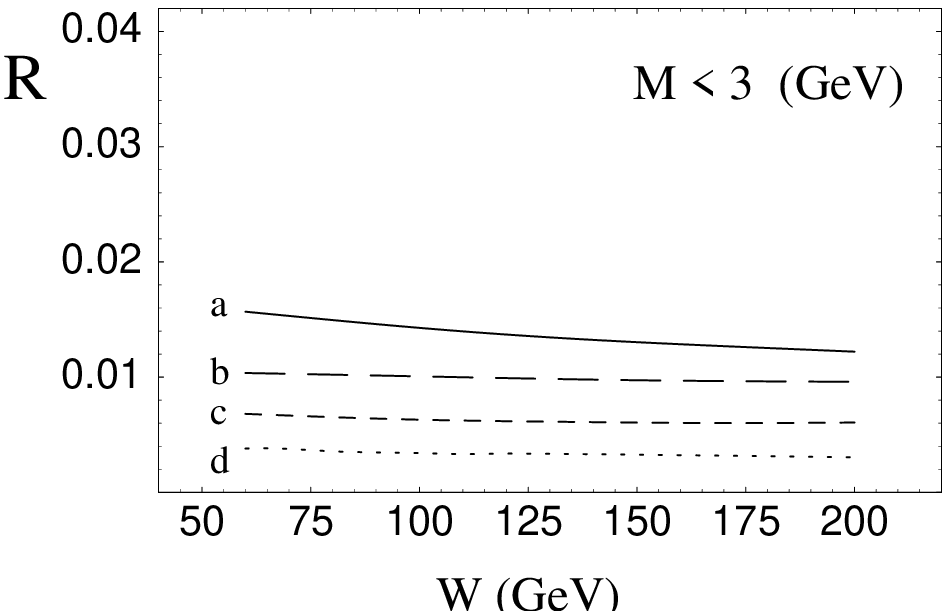,width=54mm, height=45mm}&
\epsfig{file=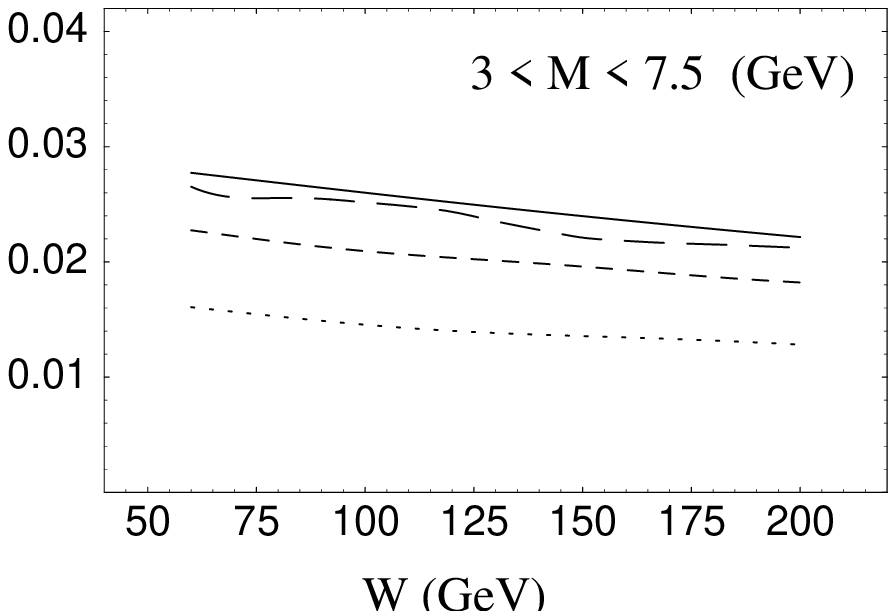,width=50mm, height=45mm}&
 \epsfig{file=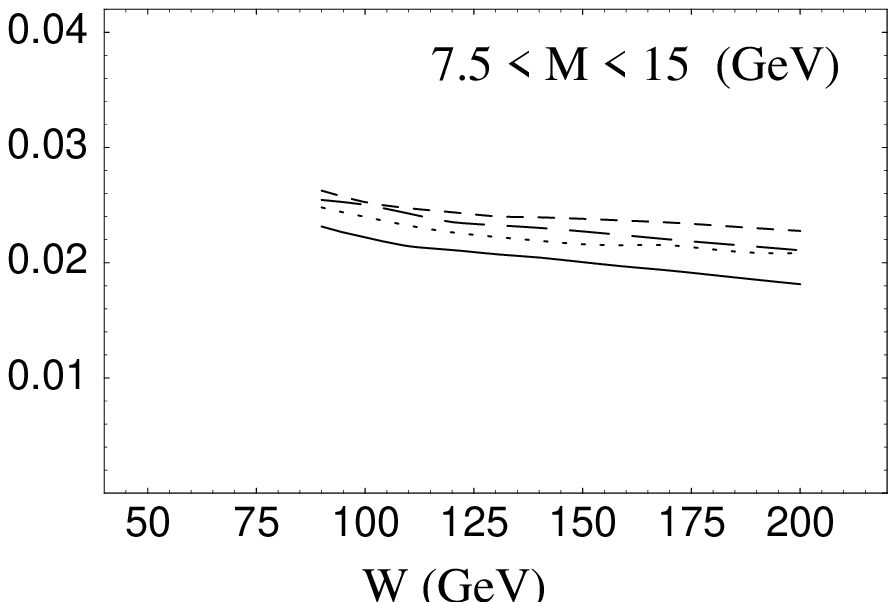,width=50mm, height=45mm}\\
\end{tabular}
  \caption[]{\it The  ratio $\si_{diff}/\si_{tot}$ as a function of $W$. a   -
 $Q^2=8\,\,GeV^2$, b -  $Q^2=14\,\,GeV^2$, c  - $Q^2=27\,\,GeV^2$, 
and d -  $Q^2=60\,\,GeV^2$. }
\label{ra}
\end{figure}

$\si_{diff}(x,x_0,Q^2)$ is the cross section for the diffractive production
of all masses below given $M^2=Q^2(x_0-x)/x$. Hence 
 the result for a mass bin can be obtained as a difference between two
cross sections corresponding to largest and smallest masses in the bin. 
Fig. \ref{ra} presents the $R=\si_{diff}/\si_{tot}$  which is a
 main result of this letter.
From Fig. \ref{ra} the ratio $R$ is observed to be 
practically independent on the energy $W$. This result 
captures the main feature of the experimental data (Fig. \ref{fig1}). However
the absolute values of the ratios are not reproduced correctly. There are 
several reasons for this discrepancy which are listed below.
\begin{itemize}
\item Due to numerical limitations the $b$-dependences of both functions $N$
and $N^D$ were simplified and both total and diffractive cross sections were
underestimated. However, as was argued above the corresponding errors are
not fully canceled in the ratio. More correct treatment of the $b$-dependence
is likely to enhance the ratio at relatively small $Q^2$.
\item Both the non-linear evolution equations used in the analysis are valid at
very low $x$. Moreover, they do not incorporate the correct DGLAP kernel at
high $Q^2$. The experimental data of Fig. \ref{fig1} covers kinematic domain 
where these equations are  expected to gain corrections due to 
DGLAP kernel \cite{LGLM,talk}. 
\item The experimental data (Fig. \ref{fig1}) includes target excitations 
which are not accounted by the evolution equations. These excitations could
in principal reach up to 30\% of the diffractive production \cite{GLMDD}.
\end{itemize} 
The above sources of the uncertainty may potentially change 
the ratios significantly. Nevertheless we believe that their
 approximate energy independence would persist in any case. In our opinion,
 this independence is rather fundamental and  related to the scaling phenomena.
 We will discuss the issue in the next section.

\section{Discussion}

In this section we will argue that the energy independence of the 
$\si_{diff}/\si_{tot}$ ratio can be traced back
to the scaling property displayed by the amplitudes $N$ and $N^D$  
and to the fact that both saturation
scales depend on $x$ with the very same power $\lambda$ \cite{LL1}.

Both the amplitudes $N$ and $N^D$ were discovered to display the remarkable
scaling phenomena \cite{me,LL1}. Namely, 
\beq\label{sc}
N(r_\perp,x;b)\,=\,N(\tau;b);\,\,\,\,\,\,\,\,\,\tau\,\equiv\, r_\perp\, Q_s(x);
\,\,\,\,\,\,\,\,\,Q_s(x)\,=\,Q_{s0}\,x^{-\lambda};\,\,\,\,\,\,\,\,\,
\lambda\,=\,0.35\,\pm \,0.04\,.
\eeq
\begin{eqnarray}\label{scD}
N^D(r_\perp,x,x_0;b)\,&=&\,N(\tau^D;b)\,;\,\,\,\,\,\,\,\,\,
\tau^D\,\equiv\, r_\perp\, Q_s^D(x,x_0)\,;\nonumber \\
\,\,\,\,\,\,\,\,\,Q_s^D(x,x_0)\,&=&\,Q_{s0}^D(x_0)\,x^{-\lambda}\,;
\,\,\,\,\,\,\,\,\lambda\,=\,0.37\,\pm\, 0.04\,.
\end{eqnarray}
The function $Q_{s0}^D(x_0)$ has a very weak dependence on $x_0$. With a quite
good accuracy the scaling (\ref{sc}, \ref{scD}) was found for all $x$ below 
$x=10^{-2}$ \cite{me,LL1}.

Assuming (\ref{sc}, \ref{scD}) to be exact property, we can plug the amplitudes
into the cross section. As a result, the ratio $R$ is given by the following
expression:
\beq\label{Rsc}
R\,=\,\frac{Q_s^{D\,2}(x,x_0^h)\,f^D(Q/Q_s^D)\,-\,
Q_s^{D\,2}(x,x_0^l)\,f^D(Q/Q_s^D)}{Q_s^2(x)\,f^{tot}(Q/Q_s)}\,.
\eeq
In (\ref{Rsc}) $x_0^{h,l}$ correspond to high and low masses in a given mass 
bin. The functions $f^{tot}$ and $f^D$ are obtained as a result of the 
dipole degree of freedom integrations.  For the sake of  transparency we use the small $z$
approximation to  simplify the wave function integration \cite{MU90,WF} :
\beq \label{D1}
\int d^2 \,r_\perp \,\int dz  \, P^{\gamma^*}(z,r_\perp;Q^2)\,\,\,\rightarrow\,\,\, 
const \times \int_{4/Q^2} 
\frac{d^2\,r_\perp}{Q^2\,\,r_\perp^4}\,.
\eeq
Using \eq{D1} one can obtain the following expressions for $\sigma_{tot}$ and 
$\sigma_{diff}$:
\begin{eqnarray}
\sigma_{tot}\,\,&=&\,\,const  \times \int_{4/Q^2}
\frac{d^2\,r_\perp}{Q^2\,\,r_\perp^4}\,\int\,d^2\,b 
\,\,N(r_{\perp},x;b)\,\,=\,\,const\,\,
\tau^2\,\int_{\tau}\,\frac{d \tau'}{\tau'^3}\,\int d^2 b 
\,\,N(\tau';b)\,\,;\nonumber  \\ \label{D2}\\
\sigma_{diff}\,&=&\,const  \times \int_{4/Q^2}
\frac{d^2\,r_\perp}{Q^2\,r_\perp^4}\int d^2\,b 
\,N^D(r_{\perp},x,x_0;b)\,=\,const\,
\tau^{D\,2}\int_{\tau^D}\,\frac{d \tau'^D}{\tau'^{D\,3}}\int d^2 b
\,N^D(\tau'^D;b)\,.\nonumber
\end{eqnarray}

These equations show that the main contribution in integration over $\tau$ stems 
from 
the region of small $\tau$ (small dipole sizes). It was shown in Refs. 
\cite{Braun,LGLM,LL,Braun2,LL1,me} that both $N(r_{\perp},x;b )$ and 
$N^D(r_{\perp},x,x_0;b)$ at low $x$ display scaling properties even at short 
distances where $N \,\propto \,r^2_{\perp}\,Q^2_s(x)$ and $ N^D 
\,\propto\,\left( \,r^2_{\perp}\,Q^{D\,\,2}_s(x,x_0)\,\right)^2$.
Substituting these estimates in \eq{D2}  one can see that 
\begin{eqnarray}
\sigma_{tot} 
\,\,&\propto  &\,\,S\,\,\tau^2\,\ln(\tau)\,\,+\,\,Const\,\,;\label{D4}\\
\sigma_{diff}\, &\propto  &\,\,S\,\tau^{D\,\,2} \,\,Const\,\,,\label{D5}
\end{eqnarray}
with  $S$ standing for the target 
transverse area.
 
Consequently, the main power dependence on $x$ (or $W$) comes from the 
saturation scales $Q_s$ and $Q_s^D$, which cancels in the ratio. As a result,
at most logararithmic dependence could be expected for the ratio.

In our opinion, the scaling property is a fundamental block in explaining
the energy independence of the ratio $R$.  
So successful GW saturation model has this scaling built in \cite{GW}.
To conclude the discussion it is important to note that the scaling (\ref{sc})
was discovered in the experimental data on the structure function $F_2$ 
\cite{scaling}. Hence  any possible corrections  to our analysis mentioned 
in the previous section are unlikely to spoil  the scaling phenomena.

\section{Conclusions}

The letter presents our attempt to reproduce the experimental data 
(Fig. \ref{fig1}) on  $\si_{diff}/\si_{tot}$ ratio as a function of the produced mass.
In particular focus  is the energy independence of the ratios, which is well 
established experimentally \cite{ZEUSDATA}. 

The analysis is  carried on a basis of the non-linear evolution equations
derived for the total DIS production in Ref. \cite{BA,KO} and for the 
diffractive production in Ref. \cite{LK}. The numerical solutions of these
equations  used for the analysis were obtained in Refs. \cite{LGLM,LL1}. 

Though our results (Fig. \ref{ra}) fail to reproduce correctly the experimental
data, they successfully reproduce the desired energy independence of the 
ratios. This independence is explained by relating it to the 
scaling phenomena which are argued to be a fundamental property of DIS at 
low  $x$ starting from $x \approx 10^{-2}$. These scaling phenomena are 
found to hold approximately even at short distances ($r_\perp \ll 1/Q_s$) which
give dominant contributions to the computed cross sections.

\section*{Acknowledgments}

The authors are very thankful  to Jochen Bartels,
Krystoff Golec-Biernat and Yuri Kovchergov
 for numerous elucidating  discussions on   
diffraction production in DIS. We would like to thank Boris Blok,
 Asher Gotsman,  Uri Maor, Eran Naftali and Kirill Tuchin
 for many helpful  and encouraging discussions. We thank DESY theory group 
and Hamburg University Institute of theoretical physics  for their 
hospitality.

 The research of 
E. L. was supported in part by the BSF grant $\#$ 9800276, by GIF grant $\#$
 I-620.-22.1411444  and by
Israeli Science Foundation, founded by the Israeli Academy of Science
and Humanities.
The work of M.L. was partially supported by the Minerva Foundation and its
financial  help is gratefully acknowledged.

\end{document}